\documentclass[journal,10pt]{IEEEtran}
\usepackage{graphicx}           
\usepackage{amsfonts}
\usepackage{url}
\usepackage{times,epsfig}
\usepackage{caption}
\usepackage{mathrsfs}
\usepackage{booktabs}
\usepackage{psfrag}
\usepackage{subfigure}
\usepackage{stfloats}
\usepackage{amssymb}
\usepackage{multirow}
\usepackage[justification=centering]{caption}
\usepackage{fancyhdr}
\usepackage[lined,ruled]{algorithm2e}
\begin{document}

\pagenumbering{arabic}
\title{Towards  Secure Blockchain-enabled Internet of Vehicles: Optimizing Consensus Management Using Reputation and Contract Theory }
\author{Jiawen Kang, Zehui Xiong, Dusit Niyato, \emph{Fellow, IEEE}, Dongdong Ye, Dong In Kim, \emph{Senior Member, IEEE}, \\Jun Zhao, \emph{Member, IEEE}
\IEEEcompsocitemizethanks{
Jiawen Kang, Zehui Xiong, Dusit Niyato, and Jun Zhao are with School of Computer Science and Engineering, Nanyang Technological University, Singapore. (emails: kavinkang@ntu.edu.sg, zxiong002@e.ntu.edu.sg, dniyato@ntu.edu.sg, junzhao@ntu.edu.sg).

 Dongdong Ye is with School of Automation, Guangdong University of Technology, China. (email: dongdongye8@163.com).

Dong In Kim is with School of Information \& Communication Engineering, Sungkyunkwan University,  Korea.
(email: dikim@skku.ac.kr).
}
\thanks{}
}
\maketitle
\pagestyle{headings}

\begin{abstract}
In the Internet of Vehicles (IoV),  data sharing among vehicles is critical to improve driving safety and enhance vehicular services. To ensure security and traceability of data sharing, existing studies utilize consensus schemes as hard security solutions to establish blockchain-enabled IoV (BIoV).  However, as miners are selected from miner candidates by stake-based voting, defending against voting collusion between the candidates and compromised high-stake vehicles becomes challenging. To address the challenge, in this paper, we propose a two-stage soft security enhancement solution: (i) miner selection and (ii) block verification. In the first stage, we design a reputation-based voting scheme to ensure secure miner selection. This scheme evaluates  candidates' reputation using  both historical interactions and recommended opinions from other vehicles. The  candidates  with high reputation are selected to be active miners and standby  miners. In the second stage, to prevent internal collusion among active miners, a newly generated block is  further verified and audited by standby miners. To incentivize the participation of the standby miners in block verification,  we adopt the contract theory to model the interactions between active miners and standby  miners, where block verification security and delay are taken into consideration. Numerical results based on a real-world dataset confirm the security and efficiency of our schemes for data sharing in BIoV.
\end{abstract}


\section{Introduction}

\subsection{Background and Motivations}
With the rapid development of automobile industry and the Internet of Things, vehicles generate a huge amount and diverse types of data through advanced on-board devices. Vehicles collect and share data to improve  driving safety and achieve better service quality~\cite{yang2018blockchain}. 
However, there exist significant security and privacy challenges for data sharing in IoV. 
On the one hand, vehicles may  not be willing to upload data to infrastructures, e.g., through road-side units,  with a centralized management architecture because of the  concern on a single point of failure and personal data manipulation. On the other hand, although Peer-to-Peer (P2P) data sharing among the vehicles can solve the issues of the centralized management architecture, it is facing with the problems of data access without authorization and security protection in a decentralized architecture. These challenges adversely affect  the circulation of vehicle  data, even forming data `island', and thus hinder the future development of IoV \cite{yue2017big}.

Recently, integrating blockchain technology with IoV has attracted increasing attention of researchers and developers because of decentralization, anonymity, and trust characteristics of blockchain.  A secure, trusted, and decentralized intelligent transport  ecosystem  is established by blockchain to solve vehicle data sharing problems \cite{yue2017big,7795984}. The authors in \cite{yang2018blockchain} proposed a decentralized trust management system for vehicle data credibility assessment using  blockchain with joint Proof-of-Work (PoW) and Proof-of-Stake (PoS) consensus schemes. Vehicle manufacturers Volkswagen \cite{volk} and Ford \cite{ford}  have  applied for patents that enable secure inter-vehicle communication through blockchain technologies.   An intelligent vehicle-trust point mechanism using proof-of-driving-based blockchain is presented to support secure communications and data sharing among vehicles \cite{SinghK17aa,DBLP}.  Li~\emph{et~al.}~\cite{li2018creditcoin} proposed a privacy-preserving incentive announcement network based on public blockchain. The Byzantine fault tolerance algorithm is adopted to incentivize vehicles to share traffic information. Nevertheless, there exists exorbitant cost to establish a blockchain  in resource-limited vehicles using computation-intensive PoW or unfair stake-based PoS \cite{wang2018survey}. Existing research attempts cannot neatly address the P2P data sharing problem among vehicles in IoV.


 In this paper, we utilize high-efficiency Delegated Proof-of-Stake (DPoS) consensus scheme as a hard security solution to develop a secure P2P data sharing system for IoV. Previous study has demonstrated that a DPoS scheme  is particularly suitable and  practical for IoV  \cite{yuan2016towards}, which performs the consensus process on pre-selected miners with moderate cost \cite{7935397}.  RoadSide Units (RSUs) as edge computing infrastructures, which are widely deployed over the whole road networks and easily reachable by vehicles,   can be the miners because of having sufficient computation and storage resources \cite{yang2018blockchain,michelin2018speedychain,lasla2018efficient}.
 These miners  play significant roles to publicly audit and store vehicle data and data sharing records in blockchain-enabled IoV (BIoV).
 Traditionally, miners in DPoS schemes are selected by stake-based voting. Note that the vehicles with stakes act as stakeholders in BIoV \cite{dpos}. The stakeholders with more stake have higher voting power. However, this approach suffers from the following collusion attacks  in BIoV:
\begin{itemize}
\item {\textbf{Miner Voting Collusion:}} Malicious RSUs collude with compromised high-stake stakeholders to be voted as miners. These malicious miners may falsely modify or discard transaction data during its mining process.  Although the malicious miners can be voted out of the BIoV by the majority of well-behaved  stakeholders in the next voting round, the stakeholders may not  participate in all the voting rounds. Thus, some malicious miners cannot be removed in a timely fashion, which enables the malicious miners to launch attacks to damage the system continuously \cite{liu2011a,voting}.
\item {\textbf{Block Verification Collusion:}} Malicious miners may internally collude with other miners to generate false results in the block verification stage, even to launch double-spending attack, which is also challenging~\cite{wang2018survey,minercollusion}.
\end{itemize}
Therefore,  it is necessary to design an enhanced DPoS consensus scheme with secure miner selection and block verification to defend against the collusion attacks in BIoV \cite{wang2018survey}.

%
\subsection{Solutions and Contributions}
Reputation is defined as the rating of an entity's trustworthiness by others  based on its past behaviors \cite{yang2018blockchain,huang2017distributed,liu2011a}. Similar to existing studies, we utilize reputation as a fair metric to propose a soft security solution for enhancing DPoS schemes through two stages: (i) secure miner selection, and (ii) reliable block verification. A reputation management scheme established on blockchain technologies is proposed for the miner selection. Miner candidates with high reputation are selected to form a miner group including active miners and standby miners, e.g., 21 active miners and 150 standby miners in Enterprise Operation System (EOS) \cite{eos1}. Each vehicle has its reputation opinion on an interacting miner candidate through a subjective logic model that combines recommended opinions from other vehicles and its own opinions based on  historical interactions into an accurate reputation opinion \cite{oren2007subjective}. All the reputation opinions of vehicles on the candidates are recorded as reliable and tamper-proof reputation records in transparent blockchain for reputation calculation.

Moreover, for secure block verification, blocks generated by active miners can be further verified and audited by standby miners to prevent internal collusion among active miners \cite{kangwcl}. Here, the active miners take turn to act as the block manager to generate and distribute unverified blocks. To incentivize the standby miners to participate in the block verification, we utilize contract theory to model interactions among the block manager and  miners to prevent collusion attacks. The block manager works as a contract designer. Meanwhile, the  miners including active miners and standby miners are followers to finish block verification for obtaining a part of transaction fee according to verification contribution \cite{kangwcl}. 

 The main contributions of this paper are summarized as follows.
\begin{itemize}
\item  We propose an enhanced DPoS consensus scheme with two-stage soft security solution for secure vehicle data sharing in BIoV.
 \item In the miner selection stage, we  introduce a  secure and efficient reputation management scheme by using a multi-weight subjective logic model. Miner are selected by reputation-based voting  for decreasing collusion between stakeholders with a lot of stake and miner candidates.
\item In the block verification stage, high-reputation standby miners are incentivized to participate in block verification using contract theory for preventing internal collusion among active miners.

\end{itemize}

The rest of this paper is organized as follows.  We present the system model and the enhanced DPoS consensus scheme with detailed steps for secure P2P vehicle data sharing  in Section II. We illustrate the secure  reputation management scheme by using the multi-weight subjective logic model in Section III. The incentive mechanism for secure block verification using contract theory is proposed in Section IV, followed by optimal contract  designing in Section V.  We illustrate numerical results in Section VI. Section VII concludes the paper.

\begin{figure}[t]\centering
  \includegraphics[width=0.5\textwidth]{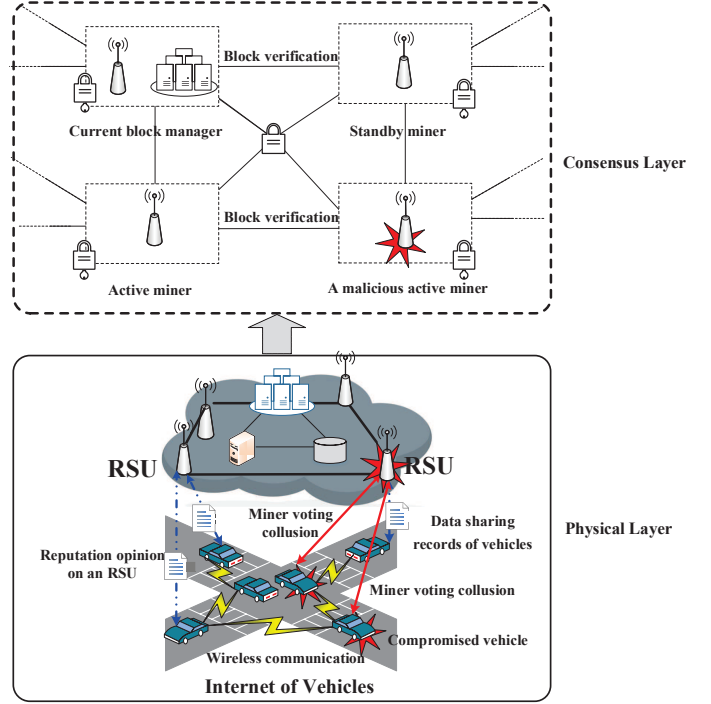}
  \caption{The system model for blockchain-based IoV.}
   \label{systemmodel}
\end{figure}
\begin{figure*}[t]\centering
  \includegraphics[width=1\textwidth]{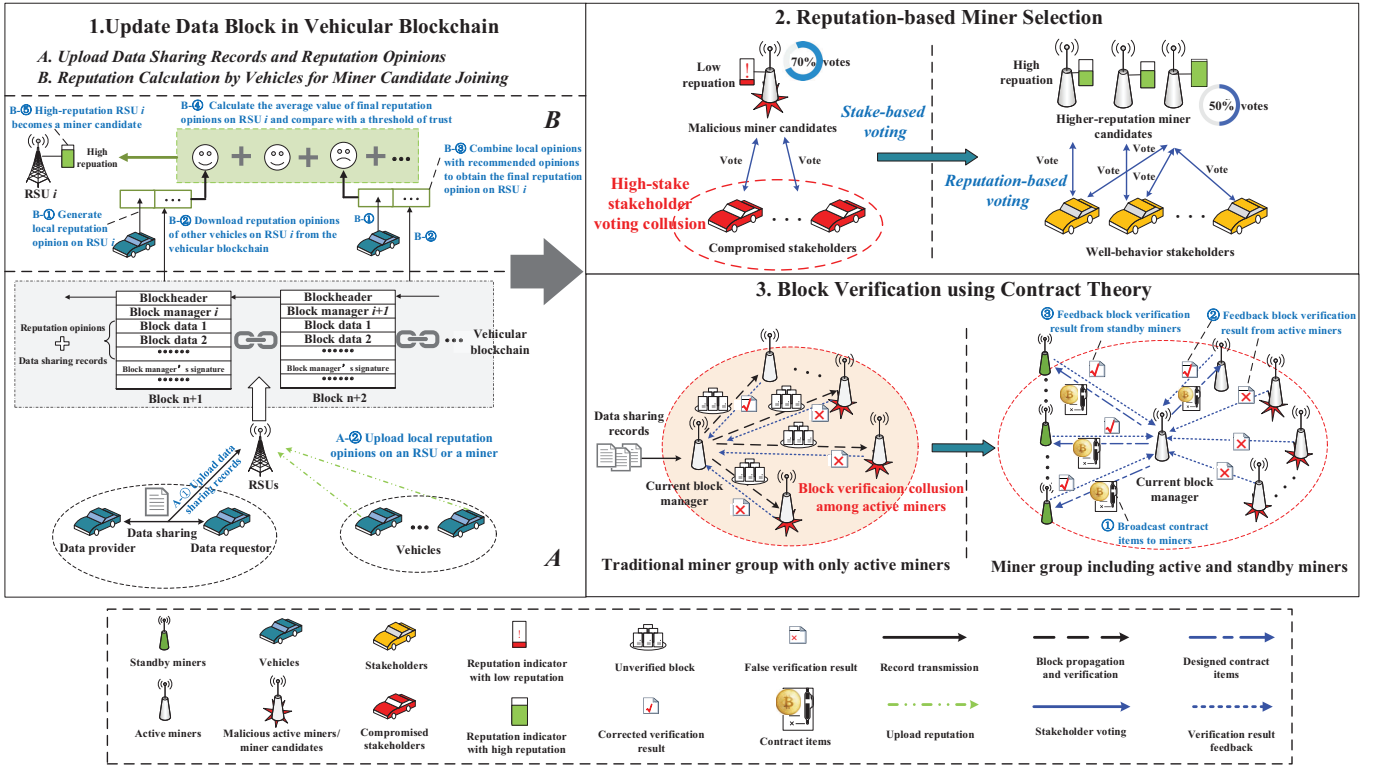}
  \caption{The enhanced DPoS consensus scheme for blockchain-based IoV.}
   \label{framework}
\end{figure*}

\section{System Model and the  Enhanced DPoS Algorithm}
\subsection{System Model}
As shown in Fig.~\ref{systemmodel}, vehicles equipped with on-board units and advanced communication devices can access vehicular services by communicating with nearby RSUs in BIoV.  The on-board units can perform simple computation, collect local data from sensing devices, and upload the data to the RSUs. Vehicles act as data collectors and share their own data with  data requesters through wireless communication. Next,  the vehicles upload their data sharing records as ``transactions" to nearby RSUs.  RSUs are deployed along roads to ensure that the vehicles are able to communicate with other vehicles and miners in a timely fashion \cite{yang2018blockchain,michelin2018speedychain,lasla2018efficient}. Unlike traditional DPoS schemes that miners are selected by stake-based voting,   RSUs with high reputation are selected as miners, whose reputation values are  calculated by a multi-weight subjective logic model. More details about  the model are given in Section III. The data collectors share data with each other and obtain a reward from data requesters. Next, the data collectors upload data sharing records to active miners, and the miners execute the consensus process of our enhanced DPoS consensus scheme. Finally, the vehicle's data sharing records are stored as block data and added into a blockchain, named {\em vehicular blockchain}, for achieving efficient proof of  presence of the data sharing.

The vehicular blockchain is also a public ledger that records vehicles' reputation opinions for RSUs and miners into the block data. These reputation opinions are persistent and transparent evidence when disputes and destruction occur \cite{lu2018bars}.  Vehicles assess both  RSUs during vehicular services and   active miners  in the consensus process. The vehicles also download the existing reputation opinions about these entities in vehicular blockchain as recommended opinions. Then, vehicles generate their reputation opinions through combining their own assessments with the recommended opinions, and upload these new opinions with digital signatures to new  active miners through nearby RSUs \cite{yang2018blockchain}. The miners perform the consensus process similar to that in data sharing.  All the vehicles can obtain the latest RSUs' reputation after the reputation opinions being added into the vehicular blockchain.  The system can calculate the average reputation of RSUs according to the reputation opinions in the vehicular blockchain, which is an important metric for the miner selection in the next round of the consensus process \cite{lu2018bars}.

\subsection{Adversary Model for DPoS Consensus Process}

In traditional DPoS consensus schemes,  miners are selected from miner candidates according to stake-based voting among stakeholders, i.e., vehicles with stake. In BIoV, as RSUs acting as miner candidates may be distributed along the road without  sufficient security protection, they are semi-trusted and may be vulnerable to be directly compromised by attackers \cite{yang2018blockchain,7721742}.  Both stakeholders and miner candidates are vulnerable to arbitrary manipulation by plutocrats \cite{voting}, and  become compromised  stakeholders and malicious miner candidates.
The plutocrats, i.e., attackers, can launch voting collusion that compromises some high-stake stakeholders with greater voting power, and ask the compromised stakeholders  to vote some certain miner candidates.
Moreover, compromised vehicles in BIoV can generate and upload fake reputation opinions to an RSU in order to increase or decrease the reputation of the target RSU \cite{yang2018blockchain}.  
Due to the overwhelming cost, we consider that the attackers cannot compromise the majority of vehicles \cite{lu2018bars}. Only a small subset of  vehicles can be compromised during a short period of time in BIoV \cite{yang2018blockchain}, i.e., due to high mobility of vehicles.

\subsection{The Enhanced DPoS Scheme for Blockchain-based IoV}
As depicted in Fig.~\ref{framework}, there are mainly three parts in the enhanced DPoS consensus scheme for secure P2P vehicle data sharing: (i) updating block data (data sharing records and reputation opinions from vehicles) and miner candidates joining, (ii) reputation-based voting for miner selection and (iii) secure block verification using contract theory. More details about steps of the proposed parts are given in the subsequent discussions.

\emph{\textbf{Step 1:}} \emph{System Initialization:}
  In vehicular blockchain, elliptic curve digital signature algorithm and asymmetric cryptography are adopted for system initialization. Every entity becomes  legitimate after passing identity authentication by a global Trust Authority (TA), e.g., a government department of transportation\footnote{Note that the TA is responsible for identity authorization, certificate issuance and access control of entities before running vehicular blockchain. That is, the TA does not affect the decentralization of the vehicular blockchain \cite{lu2018bars}.}. Each legitimate entity obtains its public \& private keys and the corresponding certificates for information encryption and decryption \cite{7935397}. An RSU that wants to be a miner candidate first submits its identity-related information to the TA. As shown in Fig. 2, the TA verifies the validity of the RSU by calculating its average reputation according to stored reputation opinions from vehicles in the vehicular blockchain. Only if the average reputation of this RSU is higher than a threshold of trust, the RSU can become a miner candidate. The threshold can be set according to different security-level requirements \cite{huang2017distributed}, which is explained in Section VI-B.

\emph{\textbf{Step 2:}} \emph{Miner candidate joining:}  Each miner candidate  submits  a deposit of stake to an account under public supervision after being a miner candidate. This deposit will be confiscated by the vehicular blockchain system if the candidate behaves maliciously and causes damage during the consensus process, e.g., failing to produce a block in its time slot \cite{eos1,8400278}. 

\emph{\textbf{Step 3:}} \emph{Reputation calculation:} As shown in Fig.~\ref{framework}, stakeholders can calculate all miner candidates' reputation by using a subjective logic model, which is based on  historical interactions with the miner candidates and recommended opinions from other vehicles. The subjective logic model takes three  weights about the historical interactions into consideration to form the local opinion on each miner candidate. The latest recommended opinions can be downloaded from the vehicular blockchain. Thus each stakeholder combines its local opinion with the recommended opinions to obtain a final reputation opinion on every miner candidate.   More details about the reputation calculation are presented  in Section III. 

\emph{\textbf{Step 4:}} \emph{Miner selection:} According to the final reputation opinions calculated by  Step~3, as shown in Fig.~\ref{framework}, each stakeholder votes for  $y$  candidates as the miners according to its ranking of the final reputation opinions for the candidates. Unlike traditional DPoS schemes, all the stakeholders have the same weight in miner voting (same voting power) even though some stakeholders owning larger stake.  The top $k$ miner candidates with the highest reputation are selected to be active miners and ($y-k$) miner candidates can be standby miners. The active miners and standby miners form a miner group in vehicular blockchain. Here $y<k$, and $k$ is an odd integer, such as 21 in EoS and 101 in Bitshares \cite{eos1}.

\emph{\textbf{Step 5:}} \emph{Block manager generation:} In line with traditional DPoS schemes,  each of the $k$ active miners takes turn to act  as the block manager during $k$  time slots of  the consensus process. Similar to that in traditional DPoS consensus schemes, every active miner  plays the role of the block manager to perform block generation, broadcasting, verification and management in its time slot.

\emph{\textbf{Step 6:}} \emph{Consensus process:} As shown in Fig.~\ref{framework}, in a time slot, the block manager first generates an unverified block, and broadcasts this block  to other active miners for block verification. However, due to the limited number of active miners, malicious active miners may launch the block verification collusion attack to generate false  block verification results.
In the block verification stage, the more verifiers result in a more secure blockchain network \cite{kangwcl}. Therefore, to defend this attack and further enhance security performance of the proposed DPoS consensus scheme, more verifiers are motivated and incentivized to participate in the block verification instead of only active miners finishing the verification.  In other words, the miners  including active miners and standby miners can act as verifiers and join the block verification process, especially the high-reputation miners, which can prevent the block verification collusion among the active miners. As such,  we then design an incentive mechanism by using contract theory to encourage high-reputation miners to participate in the block verification. In the incentive mechanism, the active miner acts as the block manager and the contract designer to broadcast contract items to miners. Meanwhile, the miners choose and sign their  best contract items.  More details about the block verification  using contract theory are described in Section IV.

In Fig.~\ref{framework}, for mutual supervision and verification, high-reputation miners locally audit the data block and broadcast their audit results with their signatures to each other. After receiving the audit results, each miner compares its result with those of other miners and sends a reply as a feedback to the block manager. This reply consists of the miner's audit result, comparison result, signatures, and records of received audit results. The  block manager analyzes the received replies from miners. If more than two third of the miners agree on the data block, the  block manager will send the records including the current audited data block and the corresponding signature to all of the miners for storage. Next, this block is stored in the vehicular blockchain. The  block manager is rewarded with cryptocurrency, and the other miners participating in block verification will receive a part of the transaction fee.  
After $k$ time slots, the group of miners and their categories, i.e., active or standby miners,  will be updated and shuffled through new miner selection.

\emph{\textbf{Step 7:}} \emph{Reputation updating:} After each round of the consensus process, vehicles download and check new data block related to their data sharing records or reputation opinions in the vehicular blockchain. If the data is correct, the vehicles will update  their reputation opinions for these miners and upload their opinions to new miners of the next round of consensus process.  The miners perform consensus process in Step 6 to add valid reputation values into the vehicular blockchain.

Note that traditional DPoS consensus schemes mainly include the following steps: miner selection, block mining and generation, and block verification. The proposed DPoS consensus scheme only enhances the miner selection step and block verification step for secure BIoV, while the block mining and generation steps are the same as those in traditional DPoS schemes. Therefore the enhanced steps are compatible with traditional DPoS schemes.

\section{Efficient Reputation Calcualtion Using Subjective Logic Model}
If a positive interaction between vehicles and RSUs/miners occurs,  the vehicles will generate a positive rating for the RSUs/miners. Consequently,  the vehicle's local reputation opinion on the RSUs/miners is increased. The positive interaction means that the vehicles believe that the services provided by RSUs is relevant and useful or the new data block  generated by a miner is true. Note that the miner candidates with high reputation acting as miners can ensure a secure and reliable consensus process. On the contrary, some compromised vehicles may  generate   fake rating because of collusion with malicious RSUs or selfish purpose.  More false ratings cause more negative effects on miner selection in the proposed DPoS scheme, thus resulting in unreliable and insecure BIoV.  Therefore, it is necessary to design a secure and efficient reputation management scheme of RSUs, and also to defend against the collusion between RSUs and vehicles. Vehicles choose their own best miner candidates as the miners according to reputation  calculation \cite{yang2016architecture}.  A multi-weight subjective logic model for reputation calculation is proposed  in this section.

Subjective logic is utilized to formulate individual evaluation of reputation based on historical interactions and recommended opinions. It is a framework for probabilistic information fusion operated on subjective beliefs about the world. The subjective logic utilizes the term ``opinion'' to denote the representation of a subjective belief, and models positive, negative statements and uncertainty. It also offers a wide range of logical operators to combine and relate different opinions \cite{huang2017distributed}.
In this paper, each vehicle (stakeholder) calculates reputation opinion taking all the recommended opinions into consideration. Due to the limited number of compromised vehicles, the false recommended opinions from the compromised vehicles have less effect on reputation calculation using subjective logic model since most vehicles are well-behaved and reliable. 

\subsection{Local Opinions for Subjective Logic}
Considering a  vehicle $V_i$ and an RSU $R_j$, the vehicle may interact with the RSU during driving, e.g., crowdsensing or vehicle data sharing. The trustworthiness (i.e., local opinion) of $V_i$ to  $R_j$ in the subjective logic can be formally described as a local opinion vector ${\omega _{i \to j}} := \{ {b_{i \to j}},{d_{i \to j}},{u_{i \to j}}\}$,
where ${b_{i \to j}},{d_{i \to j}},$ and ${u_{i \to j}}$ represent the belief, distrust, and uncertainty, respectively. We consider that all of the vehicles have the same evaluation criteria to generate local opinions. Here, $ {b_{i \to j}},{d_{i \to j}},{u_{i \to j}}\in [0,1]$ and ${b_{i \to j}} + {d_{i \to j}} + {u_{i \to j}} = 1$. According to the subjective logic model \cite{oren2007subjective,huang2017distributed}, we have
\begin{equation}
\left\{ {\begin{array}{*{20}{c}}
{{b_{i \to j}} = (1 - {u_{i \to j}})\frac{\alpha }{{\alpha  + \beta }},}\\
{{d_{i \to j}} = (1 - {u_{i \to j}})\frac{\beta }{{\alpha  + \beta }},}\\
{{u_{i \to j}} = 1 - {s_{i \to j}}.}
\end{array}} \right.
\end{equation}
$\alpha$ is the number of positive interactions and  $\beta$ is the number of negative interactions. The communication quality  $s_{i \to j}$ of a link  between vehicles $i$ and $j$, i.e., the successful transmission probability of data packets, determines the uncertainty of local opinion vector  $u_{i \to j}$  \cite{huang2017distributed}.
According to $\omega _{i \to j}$, the reputation value $T_{i \to j}$ represents the expected belief of vehicle $V_i$ that RSU $R_j$ is trusted and behaves normally during consensus process, which is denoted by
\begin{equation}
{T_{i \to j}} = {b_{i \to j}} + \gamma {u_{i \to j}}. \label{eq3}
\end{equation}
Here, $0 \leq \gamma \leq 1$ is the given constant indicating an effect level of the uncertainty for reputation \cite{oren2007subjective}.

\subsection{Multi-weight Local Opinions for Subjective Logic}
Local opinions using the subjective logic model are affected by different factors. Traditional subjective logic is evolved toward multi-weight subjective logic when considering weighting operations. Similar to \cite{huang2017distributed}, we consider the following weights to formulate local opinions.

\begin{itemize}
\item{\emph{Interaction Frequency}:} It is known that  the higher interaction frequency means that vehicle  $V_i$ has more prior knowledge about RSU $R_j$. The interaction frequency between $V_i$ and $R_j$ is the ratio of the number of times that $V_i$ interacts with $R_j$ to the average  number of times that $V_i$ interacts with other RSUs during a time window $T$, i.e.,
    \begin{equation}
    {\rm {IF}_{i \to j}}= \frac{{{N_{i \to j}}}}{{\overline {{N_i}} }},
    \end{equation}
    where ${N_{i \to j}} = ({\alpha _i} + {\beta _i}),$ and $\overline {{N_i}}  = \frac{1}{{\left|S\right|}}\sum\limits_{s \in S} {{N_{i \to s}}}$. $S$ is the set of all of the RSUs (denoted as $RSU_s$) interacting with $V_i$ during the time window. The higher interaction frequency leads to higher reputation.
\item{\emph{Interaction Timeliness}:}
In BIoV, a vehicle is not always trusted and reliable. Both the trustfulness and reputation of $V_i$ to $R_j$ are changing over time. The recent interactions have higher impact on the local opinion of $V_i$ to $R_j$. The time scale of recent interactions and past interactions is defined by $t_{recent}$, e.g., three days. The recent interactions and past interactions have different weights on the local opinions of vehicles. The parameter $\zeta$ represents the weight of recent interactions, and $\sigma$ represents the weight of past interactions. $\zeta + \sigma=1, \zeta >\sigma$.
\item{\emph{Interaction Effects}:}
 Note that positive interactions increase  RSUs' reputation and negative interactions decrease the reputation of RSUs. Therefore, the negative interactions have a higher weight on the local opinions of vehicles than that of the positive interactions. Here, the weight of positive  interactions  is $\theta$, and the weight of negative interactions is $\tau$, where $\theta  + \tau  = 1, \theta < \tau.$
The weights of interaction timeliness and  interaction effects are combined together to form a new interaction frequency as follows:
\begin{equation}
\left\{ {\begin{array}{*{20}{c}}
{{\alpha _i} = \zeta \theta \alpha _1^i + \sigma \theta \alpha _2^i,}\\
{{\beta _i} = \zeta \tau \beta _1^i + \sigma \tau \beta _2^i.}
\end{array}} \right.
\end{equation}
The positive and negative recent interactions are $\alpha _1^i$ and $\beta _1^i$ when the current time $t$ satisfies $t \le {t_{recent}}$, respectively. When $t > {t_{recent}}$, the positive and negative past interactions are $\alpha _2^i$ and $\beta _2^i$, respectively.
Therefore, the interaction frequency between two vehicles is updated as follows:
\begin{equation}
{\rm {IF}_{i \to j}} = \frac{{{N_{i \to j}}}}{{\overline {{N_i}} }} = \frac{{\theta (\zeta \alpha _1^i + \sigma \alpha _2^i) + \tau (\zeta \beta _1^i + \sigma \beta _2^i)}}{{\frac{1}{{\left| S\right|}}\sum\limits_{s \in S} {{N_{i \to s}}} }}.
\end{equation}

  Therefore, the overall weight of reputation for local opinions is ${\delta _{i \to j}} = {\rho _i}* {\rm {IF}_{i \to j}},$ where ${0 \leq \rho _i \leq 1}$ is pre-defined parameter.
 \end{itemize}
\subsection{Recommended Opinions for Subjective Logic}
After being weighted, the recommended opinions are combined into a common opinion in the form of $\omega _{x \to j}^{rec} := \{ b_{x \to j}^{rec},d_{x \to j}^{rec},u_{x \to j}^{rec}\}$. Here,
\begin{equation}
\left\{ \begin{array}{l}
b_{x \to j}^{rec} = \frac{1}{{\sum\limits_{x \in X} {{\delta _{x \to j}}} }}\sum\limits_{x \in X} {{\delta _{x \to j}}} {b_{x \to j}},\\
d_{x \to j}^{rec} = \frac{1}{{\sum\limits_{x \in X} {{\delta _{x \to j}}} }}\sum\limits_{x \in X} {{\delta _{x \to j}}} {d_{x \to j}},\\
u_{x \to j}^{rec} = \frac{1}{{\sum\limits_{x \in X} {{\delta _{x \to j}}} }}\sum\limits_{x \in X} {{\delta _{x \to j}}} {u_{x \to j}},
\end{array} \right.
\end{equation}
where $x \in X$ is  a set of recommenders that are other vehicles had interacted with $R_j$.
Thus, the subjective opinions from different recommenders are combined into one single opinion, which is called  the recommended opinion according to each opinion's weight  \cite{liu2011a}.

\subsection{Combining Local Opinions with Recommended Opinions}
After obtaining ratings of $R_j$ from  other vehicles, a particular vehicle has a subjective opinion (i.e., local opinion) on each vehicle based on its interaction history. This local opinion should still be considered while forming the final reputation opinion to avoid cheating \cite{liu2011a}.  The final reputation opinion of $V_i$ to $R_j$  is formed as  $\omega _{x \to j}^{final} := \{ b_{x \to j}^{final},d_{x \to j}^{final},u_{x \to j}^{final}\}$, where $b_{i \to j}^{final}$,  $d_{i \to j}^{final}$ and $u_{i \to j}^{final}$ are respectively calculated as follows \cite{huang2017distributed}:
\begin{equation}
\left\{ \begin{array}{l}
b_{i \to j}^{final} = \frac{{{b_{i \to j}}u_{x \to j}^{rec} + b_{x \to j}^{rec}{u_{i \to j}}}}{{{u_{i \to j}} + u_{x \to j}^{rec} - u_{x \to j}^{rec}{u_{i \to j}}}},\\[9pt]
d_{i \to j}^{final} = \frac{{{d_{i \to j}}u_{x \to j}^{rec} + d_{x \to j}^{rec}{u_{i \to j}}}}{{{u_{i \to j}} + u_{x \to j}^{rec} - u_{x \to j}^{rec}{u_{i \to j}}}},\\[9pt]
u_{i \to j}^{final} = \frac{{u_{x \to j}^{rec}{u_{i \to j}}}}{{{u_{i \to j}} + u_{x \to j}^{rec} - u_{x \to j}^{rec}{u_{i \to j}}}}.
\end{array} \right.
\end{equation}

Similar to Eqn.~(\ref{eq3}), the final reputation opinion of $V_i$ to $R_j$  is
\begin{equation}
T_{i \to j}^{final} = b_{i \to j}^{final} + \gamma u_{i \to j}^{final}.
\label{eq15}
\end{equation}
The final reputation opinions can be used in different steps of the proposed DPoS scheme. For Step 2 and Step 7 in Section II-C,  after obtaining the final reputation opinion on an RSU, vehicles will upload and store their final reputation opinions as recommended opinions for other vehicles (stakeholders) in the vehicular blockchain. For  Step 3 and Step 4 in Section II-C, stakeholders vote high-reputation miner candidates according to the reputation opinions.

\section{Incentive Mechanism for Secure Block Verification Using Contract Theory}
After selecting high-reputation miner candidates as active miners by using the multi-weight subjective logic model, there still exists a potential block verification collusion attack in the vehicular blockchain.
In this section, for secure block verification, we aim to design an incentive mechanism to motivate more miners (both active miners and standby miners) to participate in the block verification. Every block manager will offer a part of  the transaction fee as a reward to verifiers that participate in block verification and accomplish the tasks in time. Nevertheless, to do so, there are issues for the block manager in every consensus process. Firstly, the block manager does not have prior knowledge about which miners would like to participate in verification. Secondly, it does not have an accurate reputation value of a verifier. Thirdly, it does not know the amount of resource that each verifier would contribute. The information asymmetry between the block manager and verifiers may incur too much cost for the block manager to give an incentive to the verifiers. Thus, the best strategy for the block manager is  to design an incentive mechanism that can reduce the impact of information asymmetry. Moreover, the verifiers that contribute more should be rewarded more. Thus, we adopt contract theory \cite{zeng2018icc} in designing the incentive mechanism.
%

In the $k$th block verification, consider a monopoly market consisting of a block manager acting as the task publisher and a set of verifiers $\mathbb{M}=\{M_1,\ldots, M_m\}$ including active miners and standby miners.  Verifiers are willing to contribute different computation resources $C=\{c_1^k,\ldots,c_m^k\}$, i.e., CPU cycles per unit time to execute the block verification.   $I_k$  and $O_k$ are the sizes of the transmitted block before verification and the verified results, respectively \cite{zeng2018icc}. For simplicity, for all verifiers,  the values of $I_k$ and $O_k$ respectively are the same in the $k$th block verification. For a verifier $m$, the occupied CPU resource of block verification task  is $\rm{Task_{m}^{k}}$. Here, we consider that  $\rm{Task_{1}^{k}}=\rm{Task_{2}^{k}}=\cdots=\rm{Task_{m}^{k}}$. Therefore, the block verification task is denoted as a three tuple  $(\rm{Task_{m}^{k}}, I_k, O_k)$. To attract more high-reputation verifiers, we define reputation as the type of a verifier. There are  $ Q$ types, and the verifiers are sorted in an ascending order of reputation: $\theta_1<\cdots<\theta_q<\cdots< \theta_Q$, $q \in \{1,\ldots,Q\}$. The larger $\theta_q$ implies a higher reputation verifier for secure block verification among miners \cite{wang2018survey,kangwcl}.

With information asymmetry, the block manager should design specific contracts to overcome its economic loss. For different types of verifiers with different reputations, the block manager offers the verifiers  a contract $(R_q(L_q^{-1}),L_q^{-1})$, which includes a series of latency-reward bundles. Here, $L_q$ is the latency of block verification for type-$q$ verifiers and $L_q^{-1}$ is the reciprocal of $L_q$. $R_q(L_q^{-1})$ is the corresponding incentive. Note that if verifiers finish block verification faster, i.e., with smaller latency,  can be rewarded more incentive \cite{zeng2018icc}.

\subsection{Latency in Block Verification}
As mentioned in Step 6 of Section II-C, there are four steps in the block verification process for a verifier: (i) unverified block transmission from the block manager to verifiers,  (ii) local block verification, (iii) verification result broadcasting and comparison among verifiers, and (iv) verification feedback transmission from the verifiers to the manager. For a verifier $m$, the  latency  consisting of the corresponding  delays of the aforementioned steps is  defined  as follows \cite{zeng2018icc},
\begin{equation}
{{L}_{q}}(c_{m}^{k},{{I}_{k}},{{O}_{k}})=\frac{{{I}_{k}}}{r_{m}^{d}}+\frac{\rm{Task_{m}^{k}}}{c_{m}^{k}}+\psi {{I}_{k}}\left| \mathbb{M} \right|+\frac{{{O}_{k}}}{r_{m}^{u}}.
\end{equation}
 $r^u_m$ is the uplink transmission rate from the verifiers to block manager and  $r^d_m$ is the downlink transmission rate from  the block manager to the verifiers.
The transmission time of an unverified block from the block manager to the verifier is $\frac{{{I}_{k}}}{r_{m}^{d}}$. The local verification time of this block is $\frac{\rm{Task_{m}^{k}}}{c_{m}^{k}}$.  Similar to that in \cite{kangwcl, liu2017evolutionary}, the time of verification result broadcasting and comparison among verifiers is a function of the  block size $I_k$, network scale (i.e., the number of verifiers $\left| \mathbb{M} \right|$) and average verification speed of each verifiers, which is denoted as  $\psi {{I}_{k}}\left|\mathbb{M} \right|$. Here, $\psi$ is a pre-defined parameter of verification result  broadcasting and comparison, which can be obtained from statistics of previous block verification processes. The time of verification feedback is  $\frac{{{O}_{k}}}{r_{m}^{u}}$.

$r_m^u$ and $r_m^d$ can be calculated based on wireless link speed, e.g., the  Shannon capacity. Let locations of verifiers fix during block verification. We apply the Time-Division Medium Access (TDMA) technique, where the uplink and downlink use the same frequency channel  \cite{zeng2018icc}. Then, we have
\begin{equation}
r_{m}^{u}=r_{m}^{d}=B{{\log }_{2}}(1+\frac{{\varpi_{m}}|{{h}_{m}}{{|}^{2}}}{\sum\limits_{{{m}^{-}}\in \mathbb{M}\backslash \{m\}}{{\varpi_{{{m}^{-}}}}|{{h}_{{{m}^{-}}}}{{|}^{2}}}+{{N}_{0}B}}),
\end{equation}
where $B$ is the transmission bandwidth and $\varpi_m$ is the transmission power of verifier $m$. $h_m$ is the channel gain of peer-to-peer link between the verifier $m$ and the block manager or other verifiers.
$N_0$ is the one-sided power spectral density level of white Gaussian noise, and $m^-$ is an element in $\mathbb{M}$ excluding $m$.

\subsection{Profit of the Block Manager}
According to the signed contract $(R_q, L_q^{-1})$ between the block manager and type-$q$ verifier,  the profit of the block manager obtained from  type-$q$ verifier is denoted as
\begin{equation}
{{U}_{bm}}(q)=\pi[\phi_q(L_q)]-l{{R}_{q}},
\end{equation}
where $l$ is a pre-defined weight parameter about the type-$q$ verifier's incentive $R_q$.  $\pi[\phi_q(L_q)]$ is the benefit of the block manager regarding a security-latency metric $\phi_q$ for type-$q$ verifier. Intuitively, the block manager obtains a higher profit when the $\phi_q$ is bigger. Moreover,   both more high-reputation verifiers and less latency can lead to bigger $\phi_q$, i.e., $\frac{{\partial \pi (\phi_q)}}{{\phi_q}} > 0$, $\frac{{\partial \phi_q({L_q})}}{{{\theta _q}}} > 0$ and $\frac{{\partial \phi_q({L_q})}}{{{L_q}}} < 0.$ 
The more verifiers participating in block verification leads to more secure block verification stage. However, this causes larger latency since the verifiers may need to communicate with verifiers through multi-hop relays \cite{kangwcl}. Similar to that in \cite{kangwcl,chen2015decentralized}, we define a more general security-latency metric to balance the network scale and the block verification time for type-$q$ verifier,  which is expressed by
\begin{equation}
\phi_q= \left\{ {\begin{array}{*{20}{c}}
\begin{array}{l}
{e_1}{({\theta _q}|\mathbb{M}|{p_q})^{{z_1}}} - {e_2}{(\frac{{{L_q}}}{{{T_{\max }}}})^{{z_2}}},~
{\rm{if~}}0 < {L_q} <A,
\end{array}\\
{0,{\rm{    ~~~~~      ~~~~~ ~~~~~~~~~~~~    ~~~~                  }}\mbox{otherwise}.}
\end{array}} \right.
\end{equation}
Here $A=\frac{{{T_{\max }}{e_1}^{z_2^{ - 1}}{{({\theta_q}|\mathbb{M}|{p_q})}^{\frac{{{z_1}}}{{{z_2}}}}}}}{{{e_2}^{z_2^{ - 1}}}}$. $e_1>0$ and $e_2>0$ are pre-defined coefficients about the network scale and verification latency, respectively. $p_q$ is the prior probability of type-$q$, and $\sum\nolimits_{q = 1}^Q {{p_q}}=1$.
We consider that the block manager can obtain the distribution of verifier types from observations and statistics of previous behaviours of the verifiers \cite{zeng2018icc}.
$T_{max}$ denotes the maximum tolerable block verification  latency to blockchain users. $z_1\geq1$ and $z_2\geq1$ are given factors indicating the effects of  network scale and verification latency on block verification, respectively.
The goal of the block manager is to maximize its profit through block verification as follows:
\begin{equation}
\mathop {\max }\limits_{({R_q},L_q^{ - 1})} {\mkern 1mu} {U_{bm}}({q}) = \sum\limits_{q = 1}^Q {(|\mathbb{M}|{p_q})} (\pi[\phi_q(L_q)] - l{R_q}).  \label{15}
\end{equation}

\subsection{Utility of Block Verifiers}

For type-$q$ verifier, the utility function of block verification based on a signed contract is defined as
\begin{equation}
{U_q} = {\theta _q}\eta ({R_q}) - {{l'}}L_q^{ - 1},
\end{equation}
where $\eta ({R_q})$ is a monotonically increasing valuation function of type-$q$ verifier in terms of the incentive $R_q$.
$l'$ is the unit resource cost of block verification. Moreover, the valuation is zero when there is no incentive, i.e.,  $\eta (0)=0$. The higher type-$q$ verifier should have larger utility because of higher reputation in block verification. However, the verifier wants to maximize its utility through minimizing resource consumption in block verification.
Specifically, the objective of type-$q$ verifier is to maximize utility obtained by joining  block verification, expressed by
\begin{equation}
{\mathop {{\rm{max}}}\limits_{{\rm{(}}{{\rm{R}}_{\rm{q}}}{\rm{,L}}_{\rm{q}}^{{\rm{ - 1}}}{\rm{)}}} {U_q} = {\theta _q}\eta ({R_q}) - l'L_q^{ - 1},\forall q \in \left\{ {1,\ldots,Q} \right\}}. \label{17}
\end{equation}


\section{Optimal Contract Designing}
According to \cite{bolton2005contract},  to make contracts feasible, each contract item for verifiers must satisfy the following principles: (i) Individual Rationality (IR) and (ii) Incentive Compatibility (IC). IR means that each verifier will join the block verification when it receives a non-negative utility, i.e.,
\begin{equation}
{\theta _q}\eta ({R_q}) - {{l'}}L_q^{ - 1} \ge 0,\forall q \in \left\{ {1,\ldots,Q} \right\}.
\end{equation}
IC refers to that type-$q$ verifier can only receive the maximum utility  when choosing the contract designed for itself instead of all other contracts $(R_{q'}, L_{q'}^{-1})$, i.e.,
\begin{equation}
\begin{array}{l}
{\theta _q}\eta ({R_q}) - {{l'}}L_q^{ - 1} \ge {\theta _q}\eta ({R_{q'}}) - {{l'}}L_{q'}^{ - 1},\\
\forall q,q' \in \left\{ {1,\ldots,Q} \right\},q \ne q'.
\end{array}
\end{equation}

In what follows, we consider $\pi[\phi_q(L_q)]=g_1[{e_1}{({\theta _q}|\mathbb{M}|{p_q})^{{z_1}}} - {e_2}{(\frac{{{L_q}}}{{{T_{\max }}}})^{{z_2}}}]$ for ease of presentation, where $g_1$ is unit profit gain for the block manager.
Therefore, the optimization problems in (\ref{15}) and (\ref{17}) can be defined as follows:
\begin{equation}
\begin{array}{l}
\begin{array}{*{20}{l}}
{\mathop {\max }\limits_{({{\rm{R}}_{\rm{q}}},{\rm{L}}_{\rm{q}}^{{\rm{ - 1}}})} {U_{bm}} = \sum\limits_{q = 1}^Q {|\mathbb{M}|{p_q}} [{g_1}{e_1}{{({\theta _q}|\mathbb{M}|{p_q})}^{{z_1}}} - {g_1}{e_2}{{(\frac{{{L_q}}}{{{T_{\max }}}})}^{{z_2}}}}\\
{\;\;\;\;\;\;\;\;\;\;\;\;\;\;\;\;\;\;~~~~ - l{R_q}]}
\end{array}\\
{\rm{s}}.{\rm{t}}.\\
{\theta _q}\eta ({R_q}) - {{l'}}L_q^{ - 1} \ge 0, \forall q \in \left\{ {1,\ldots,Q} \right\},\\
{\theta _q}\eta ({R_q})- {{l'}}L_q^{ - 1} \ge {\theta _q}\eta ({R_{q'}}) - {{l'}}L_{q'}^{ - 1},  \forall q,q' \in \left\{ {1,\ldots,Q} \right\}, \\ q \ne q',\\
\max \{ {L_q}\}  \le {T_{\max }}, \forall q \in \left\{ {1,\ldots,Q} \right\},\\
\sum\limits_{q = 1}^Q {|\mathbb{M}|{p_q} {R_q}}  \leq {R_{\max }}, \forall q \in \left\{ {1,\ldots,Q} \right\},
\end{array}
\end{equation}
where $R_{max}$ is a given transaction fee from blockchain users.


This problem is not a convex optimization problem. However, we can find its solution by performing the following transformation.

\textbf{Lemma 1 (Monotonicity).} For contract $({R_i},{L_{i}^{-1}})$ and $({R_j},{L_{j}^{-1}})$, we have $R_i \ge R_j$ and $L_{i}^{-1} \ge L_{j}^{-1}$, if and only if ${\theta _i} \ge {\theta _j}$, $i \ne j,$ and $i,j \in \left\{ {1,\ldots,Q} \right\}$.

\textbf{Proof}: According to the IC constraints of type-$i$  verifier and type-$j$ verifier, we have
\begin{equation}
{\theta _i}\eta ({R_i}) - l'L_i^{ - 1} \ge {\theta _i}\eta ({R_j}) - l'L_j^{ - 1},\label{19}
\end{equation}
\begin{equation}
{\theta _j}\eta ({R_j}) - l'L_j^{ - 1} \ge {\theta _j}\eta ({R_i}) - l'L_i^{ - 1}.\label{20}
\end{equation}
By adding together (\ref{19}) and (\ref{20}),  we can obtain $({\theta _i} - {\theta _j})[\eta ({R_i}) - \eta ({R_j})] \ge 0.$  $\eta ({R_q}) \ge 0$ is a monotonically increasing valuation function of $R_q$.
When $\theta _i \ge \theta_j$, we can deduce that $\eta ({R_i}) - \eta ({R_j}) \ge 0$, i.e., ${R_i} \ge {R_j}$. When ${R_i} \ge {R_j}$, we have $\eta ({R_i}) - \eta ({R_j}) \ge 0$. Thus, we can deduce that $\theta _i \ge \theta_j$ must be satisfied \cite{8239591}.

\textbf{Proposition 1:}  ${R_i} \ge {R_j}$, if and only if $L_i^{ - 1} \ge L_j^{ - 1}$.

\textbf{Proof}: According to the IC constraint in (\ref{19}), we can obtain
\begin{equation}
{\theta _i}[\eta ({R_i}) - \eta ({R_j})] \ge l'(L_i^{ - 1} - L_j^{ - 1}), \label{2}
\end{equation}
\begin{equation}
{\theta _j}[\eta ({R_i}) - \eta ({R_j})] \le l'(L_i^{ - 1} - L_j^{ - 1}).  \label{3}
\end{equation}
As $L_i^{ - 1} \ge L_j^{ - 1}$, we have $\eta ({R_i}) \ge  \eta ({R_j})$ according to (\ref{2}), and thus ${R_i} \ge {R_j}$. In addition, when ${R_i} \ge {R_j}$, we can obtain $L_i^{ - 1} \ge L_j^{ - 1}$ from (\ref{3}).
\textbf{Proposition 1} indicates that an  incentive compatibility contract requires a higher payment, if verifiers have less latency in block verification.~~~~~~~~~~~~~~~~$\blacksquare$

\textbf{Lemma 2.} If the IR constraint of type-$1$ verifier is satisfied, the IR constraints of other types  will hold.

\textbf{Proof}: According to the IC constraints, $\forall i \in \{2, \dots, Q\}$, we have
\begin{equation}
{\theta _i}\eta ({R_i}) - l'L_i^{ - 1} \ge {\theta _i}\eta ({R_1}) - l'L_1^{ - 1}.\label{4}
\end{equation}
Given that $\theta_1<\cdots<\theta_i<\cdots< \theta_Q$,  we also have
\begin{equation}
 {\theta _i}\eta ({R_1}) - l'L_1^{ - 1} \ge {\theta _1}\eta ({R_1}) - l'L_1^{ - 1}.\label{5}
\end{equation}
According to (\ref{4}) and (\ref{5}), we have
\begin{equation}
 {\theta _i}\eta ({R_i}) - l'L_i^{ - 1} \ge {\theta _1}\eta ({R_1}) - l'L_1^{ - 1} \ge 0.\label{6}
\end{equation}
The (\ref{6}) indicates that with the IC condition, when the IR constraint of type-$1$ verifier is satisfied, the other IR constraints  will also hold. Therefore, the other IR constraints can be bound into the IR condition of type-$1$ verifier \cite{8239591}.~~~~~~~~~~~~~~~~~$\blacksquare$

\textbf{ Lemma 3. } By utilizing the monotonicity in \textbf{Lemma ~{1}}, the IC condition can be transformed into the Local Downward Incentive Compatibility (LDIC), which is given as follows:
\begin{equation}\label{equation} 
{{\theta _i}\eta ({R_i}) - {{l'}}L_i^{ - 1} \geq {\theta _i}\eta ({R_{i-1}}) - {{l'}}L_{i-1}^{ - 1}},\forall i \in \left\{ {2,\ldots,Q} \right\}.\\
\end{equation}

\textbf{Proof}: The IC constraints between type-$i$ and type-$j$, $j \in \{1, \dots, i-1 \}$ are defined as Downward Incentive Compatibility  (DIC), given by ${\theta _i}\eta ({R_i}) - l'L_i^{ - 1} \ge {\theta _i}\eta ({R_j}) - l'L_j^{ - 1}.$

The IC constraints between type-$i$ and type-$j$, $j \in \{i+1, \dots, Q  \}$ are defined as Upward Incentive Compatibility (UIC), given by ${\theta _i}\eta ({R_i}) - l'L_i^{ - 1} \ge {\theta _i}\eta ({R_j}) - l'L_j^{ - 1}.$

We first prove that DIC can be reduced as two adjacent types in DIC, called LDIC. Consider three continuous types of verifiers, i.e.,  $\theta_{i-1}<\theta_i< \theta_{i+1}$, $i \in \{2, \dots, Q-1\}$, we have
\begin{equation}
{\theta _{i+1}}\eta ({R_{i+1}}) - l'L_{i+1}^{ - 1} \ge {\theta _{i+1}}\eta ({R_i}) - l'L_i^{ - 1},\label{7}
\end{equation}
\begin{equation}
{\theta _{i}}\eta ({R_{i}}) - l'L_{i}^{ - 1} \ge {\theta _{i}}\eta ({R_{i-1}}) - l'L_{i-1}^{ - 1}.\label{8}
\end{equation}
According to the monotonicity, i.e.,  if ${\theta _i} \ge {\theta _j}$, then $R_i \ge R_j$, $i \ne j,$ and $i,j \in \left\{ {1,\ldots,Q} \right\}$, we have
\begin{equation}
({\theta _{i+1}} - {\theta _i})[\eta ({R_i}) - \eta ({R_{i-1}})] \ge 0,\label{9}
\end{equation}
\begin{equation}
{\theta _{i+1}}[\eta ({R_{i}}) - \eta ({R_{i-1}})]  \ge {\theta _{i}}[\eta ({R_{i}}) - \eta ({R_{i-1}})] .\label{10}
\end{equation}
Combine (\ref{8}) and (\ref{10}), we have ${\theta _{i+1}}[\eta ({R_{i}}) - \eta ({R_{i-1}})] \ge {\theta _{i}}[\eta ({R_{i}}) - \eta ({R_{i-1}})] \ge l'L_{i}^{ - 1}- l'L_{i-1}^{ - 1}$.
Thus, we have
\begin{equation}
{\theta _{i+1}}\eta ({R_{i}}) - l'L_{i}^{ - 1} \ge {\theta _{i+1}}\eta ({R_{i-1}})- l'L_{i-1}^{ - 1}. \label{11}
\end{equation}
Combine (\ref{7}) and (\ref{11}), we have
\begin{equation}
{\theta _{i+1}}\eta ({R_{i+1}}) - l'L_{i+1}^{ - 1} \ge {\theta _{i+1}}\eta ({R_{i-1}})- l'L_{i-1}^{ - 1}.\label{12}
\end{equation}
We can extend (\ref{12}) to prove that the DIC can be held until type-$1$:
\begin{equation}
\begin{array}{l}
{\theta _{i + 1}}\eta ({R_{i + 1}}) - l'L_{i + 1}^{ - 1} \ge {\theta _{i + 1}}\eta ({R_{i - 1}}) - l'L_{i - 1}^{ - 1} \ge \\
 \cdots  \ge {\theta _1}\eta ({R_1}) - l'L_1^{ - 1}, \forall~ i.
\end{array}
\end{equation}
Hence, note that with the LDIC and the monotonicity,  the DIC holds. Similarly, with the monotonicity and the Local Upward Incentive Compatibility (LUIC), the UIC can be proved to hold \cite{bolton2005contract,8239591}.

 According to \textbf{Lemmas 1}, \textbf{2}, and \textbf{3}, the optimization problem can be reformulated as follows:
\begin{equation} \label{18}
\begin{array}{l}
\mathop {\max }\limits_{({{\rm{R}}_{\rm{q}}},{\rm{L}}_{\rm{q}}^{{\rm{ - 1}}})} {U_{bm}} = \sum\limits_{q = 1}^Q {|\mathbb{M}|{p_q}} [{g_1}{e_1}{({\theta _q}|\mathbb{M}|{p_q})^{{z_1}}} - {g_1}{e_2}{(\frac{{{L_q}}}{{{T_{\max }}}})^{{z_2}}} \\~~~~~~~~~~~~~~~~~~- l{R_q}]\\
{\rm{s}}.{\rm{t}}.\\
{\theta _1}\eta ({R_1}) - l'L_1^{ - 1} = 0,\\
{\theta _q}\eta ({R_q}) - l'L_q^{ - 1} = {\theta _q}\eta ({R_{q - 1}}) - l'L_{q - 1}^{ - 1}, \forall q \in \left\{ {2,\ldots,Q} \right\},\\
\max \{ {L_q}\}  \le {T_{\max }}, \forall q \in \left\{ {1,\ldots,Q} \right\},\\
\sum\limits_{q = 1}^Q {|\mathbb{M}|{p_q} {R_q}}  \leq {R_{\max }}, \forall q \in \left\{ {1,\ldots,Q} \right\}.
\end{array}
\end{equation}
Furthermore, to simplify the  analysis without loss of generality, we define the concave function $\eta ({R_q})=R_q$.
The optimization problem in (\ref{18}) is solved sequentially. Firstly, we solve the relaxed problem in (\ref{18}) without monotonicity to obtain a solution. Secondly, we verify that whether the solution satisfies the condition of the monotonicity. We use the method of iterating the $IC$ and $IR$ constraints to obtain  $R_q$ which can be expressed as follows:
\begin{equation}\label{1} 
{R_{q}} = \frac{{{l'L_{1}^{-1} }}}{{{\theta _{1}}}} + \sum\nolimits_{k= 2}^{q} {{\Delta _{k}}},\label{22}
\end{equation}
where ${\Delta _{k}} = \frac{{{l'L_{k}^{-1} }}}{{{\theta _{k}}}} - \frac{{{l'L_{k-1}^{-1}}}}{{{\theta _{k}}}}$ and ${\Delta _{1}} = 0$.
By substituting $R_{q}$ into $\sum\limits_{q = 1}^Q {|\mathbb{M}|{p_q} {R_q}}$, we have
\begin{equation} \label{a1}
\sum\limits_{q = 1}^Q {|\mathbb{M}|{p_q}l{R_q}}=|\mathbb{M}|\sum_{q=1}^{Q}{l f_q L_q^{ - 1}},
\end{equation}
where
\begin{equation}
{f_q} = \left\{ {\begin{array}{*{20}{l}}
{\frac{{l'{p_q}}}{{{\theta _q}}} + \left( {\frac{{l'}}{{{\theta _q}}} - \frac{{l'}}{{{\theta _{q + 1}}}}} \right)\sum\limits_{i = q + 1}^Q {{p_i}},~~ if~q < Q,}\\
{\frac{{l'{p_Q}}}{{{\theta _Q}}},~~~~~~~~~~~~~~~~~~~~~~~~~~~~~if~q = Q.}
\end{array}} \right.
\end{equation}
We substitute the expression in  (\ref{a1}) into the problem in (\ref{18}) and remove all $R_{q}$, $\forall q  \in \left\{ {1,\ldots,Q} \right\}$ from the problem in (\ref{18}). The problem in (\ref{18}) is rewritten as follows:
\begin{equation}
\hspace{-10pt}\begin{array}{l}
 \mathop {\max }\limits_{({{\rm{R}}_{\rm{q}}},{\rm{L}}_{\rm{q}}^{{\rm{ - 1}}})} {U_{bm}} = \sum\limits_{q = 1}^Q {|\mathbb{M}|{p_q}} [{g_1}{e_1}{({\theta _q}|\mathbb{M}|{p_q})^{{z_1}}} - {g_1}{e_2}{(\frac{1}{{L_q^{ - 1}{T_{\max }}}})^{{z_2}}}] \\
~~~~~~~~~~~~ -|\mathbb{M}|l\sum\limits_{q = 1}^Q {{f_q}L_q^{ - 1},} \\
{\rm{s}}.{\rm{t}}.\;\;\;L_q^{ - 1} \ge \frac{1}{{{T_{\max }}}},\forall q \in \left\{ {1,\ldots,Q} \right\},\\
~~~~~~|\mathbb{M}|\sum\limits_{q = 1}^Q {f_q}L_q^{ - 1}  \leq {R_{\max }}, \forall q \in \left\{ {1,\ldots,Q} \right\}.
\end{array}  \label{23}
\end{equation}

By differentiating $U_{bm}$ with respect to $L_q^{-1}$, we have $\frac{{\partial {U_{bm}}}}{{\partial L_q^{ - 1}}} = \frac{{|\mathbb{M}|{g_1}{e_2}{z_2}{p_q}}}{{T_{\max }^{{z_2}}}}{(L_q^{ - 1})^{ - ({z_2} + 1)}} - |\mathbb{M}|l{f_q}$, and $\frac{{{\partial ^2}{U_{bm}}}}{{\partial {{(L_q^{ - 1})}^2}}} =  - \frac{{|\mathbb{M}|{g_1}{e_2}{z_2}{p_q}({z_2} + 1)}}{{T_{\max }^{{z_2}}}}{(L_q^{ - 1})^{ - ({z_2} + 2)}} < 0.$ Thus, the function $U_{bm}$ is concave. The problem defined in (\ref{23})  is a convex optimization problem because the summation of concave functions ($U_{bm}$) is still a concave function, and the constrains are affine. We can obtain the optimal latency requirement ${L_q^{-1}}^*$  and the corresponding  incentive $R_q^*$ by using convex optimization tools.
Moreover, if the types of verifiers follow uniformly distributed, the monotonicity can be automatically met  \cite{bolton2005contract,8239591}. If not,  we can use infeasible sub-sequence replacing algorithm to satisfy the final optimal latency requirement \cite{gao2011spectrum}.

Note that the proposed incentive mechanism based on contract theory can encourage efficiently high-reputation miners to join  the block verification for further improving the security of the vehicular blockchain.

\section{ Numerical Results}

\begin{figure}\centering
  \includegraphics[width=0.5\textwidth]{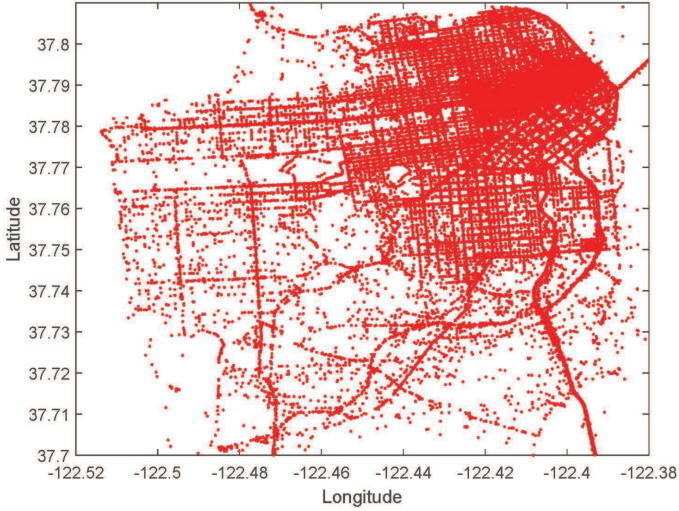}
  \caption{Spatial distribution of vehicle trace points.\\
   \small{}}\label{taix}
\end{figure}

In this section, we first evaluate the performance of the proposed Multi-Weight Subjective Logic (MWSL) scheme based on a real-world dataset of San Francisco Yellow Cab \cite{hoque2012analysis}. Next, we evaluate and compare the performance of the proposed incentive mechanism based on contract theory. 
The mobility traces of 536 taxis driving during a month are recorded in this dataset. We observe 200 taxis running in an urban area, whose latitude and longitude are from 37.7 to 37.81 and from -122.52 to -122.38, respectively. Fig.~\ref{taix} shows trace points of the 200 taxis during a month. The average time gap between two trace records is 43.34 seconds.  There are 400 RSUs (miner candidates) deployed uniformly in the observation area. The update period of RSUs' reputation is 1 minute. These miner candidates are initially classified into 10 types according to their reputation values, wherein the probability for an candidate belonging to a certain type is 0.1.
Major parameters used in the simulation are given in Table I, most of which  are adopted from \cite{huang2017distributed,zeng2018icc,8239591}.

\begin{table}[t]
\renewcommand{\arraystretch}{1.2}
\caption{Parameter Setting in the Simulation}\label{table} \centering \tabcolsep=4pt
\begin{tabular}{p{3.3cm}|p{5.2cm}}
\hline
Parameter & Setting\\
\hline
Interaction frequency between vehicles and RSUs &  [50, 200]~times/week\\
\hline
Coverage range of RSUs &  [300, 500]~m\\
\hline
Speed of vehicles &  [50, 150]~km/h\\
\hline
Weight parameters & $\theta=0.4$, $\tau=0.6$,  $\zeta=0.6$, $\sigma=0.4$, $\rho=1$ \\
\hline
 Time  scale of recent and past events $t_{recent}$   & three days \\
\hline
Rate of compromised  vehicles  & [10\%, 90\%]\\
\hline
Successful transmission probability of data packets  &  [0.6, 1]\\
\hline
Vehicle to RSU bandwidth & 20 MHz\\
\hline
Noise spectrum density & -174 dBm/Hz\\
\hline
Transmission power & [10, 23]~dBm\\
\hline
Receiver power & 14~dBm\\
\hline
Computation resource & [$10^3$, $10^6$]~CPU cycles/unit time \\
\hline
Input/output block data size & [50, 500]~KB\\
\hline
Pre-defined parameters & $g_1=1.2$, $e_1=15,~ e_2=10, z_1=2, \newline z_2=1,~ l=5,~ l'=1,~T_{max}=300~$s, \newline $R_{max}=1000,~ \psi=0.5$\\
\hline
\end{tabular}\label{table1}
\end{table}

\subsection{Performance of the proposed reputation scheme}
In the proposed MWSL scheme, vehicles calculate reputation value of miner candidates according to local opinions and recommended opinions from other vehicles. We compare our MWSL scheme with a Traditional Subjective Logic (TSL) scheme which is a typical model using a linear function to calculate reputation \cite{huang2017distributed}. More specifically, $T_{i\rightarrow j}^l=(1-\kappa)T_{ave}+\kappa T_{las}$, where $T_{ave}=b_{x\rightarrow j}^{ave}+0.5u_{x\rightarrow j}^{ave}$ and $T_{las}=b_{i\rightarrow j}^{las}+0.5u_{i\rightarrow j}^{las}$. Here $\kappa$ is the weight and is set to be 0.5. $b_{i\rightarrow j}^{ave}$ and $u_{i\rightarrow j}^{ave}$ are average values of other vehicles' $b_{i\rightarrow j}$ and  $u_{i\rightarrow j}$, respectively. $b_{i\rightarrow j}^{las}$ and $u_{i\rightarrow j}^{las}$ are the latest $b_{i\rightarrow j}$ and $u_{i\rightarrow j}$ in the local opinion of vehicle $i$ for RSU $j$.
We consider a malicious miner candidate will firstly pretend to behave well to obtain positive reputation values from vehicles in the former 5 minutes. Then, this candidate colludes with 10 compromised vehicles and begins to misbehave to 50 well-behaved vehicles randomly. These misbehaving vehicles will generate negative reputation opinions for the candidate, while the colluded vehicles  still generate positive reputation opinions for the candidate and vote it as a miner in the voting stage.

\begin{figure}[t]\centering
  \includegraphics[width=0.5\textwidth]{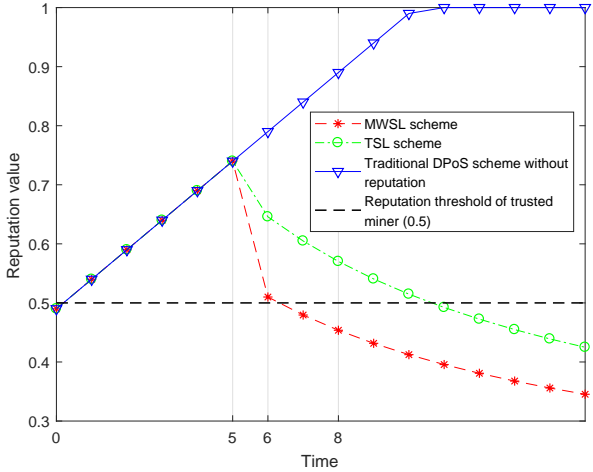}
  \caption{The reputation values of a malicious miner.}
   \label{reputation_changed1}
\end{figure}

\begin{figure}[t]\centering
  \includegraphics[width=0.5\textwidth]{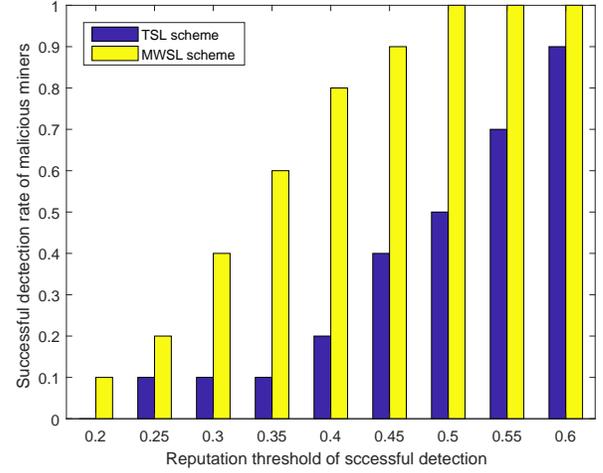}
  \caption{Detection rate of malicious miners under different threshold values of trusted miners}
   \label{detectionRate}
\end{figure}

\begin{figure}[t]\centering
  \includegraphics[width=0.5\textwidth]{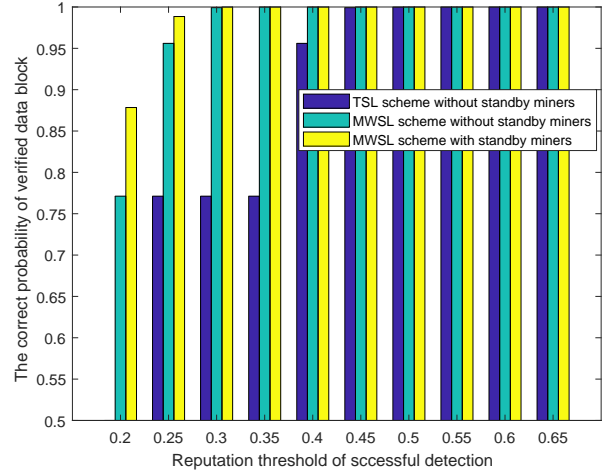}
  \caption{Probability of corrected data blocks under different threshold values of trusted miners}
   \label{correctedProbability}
\end{figure}

\begin{figure}[t]\centering
  \includegraphics[width=0.5\textwidth]{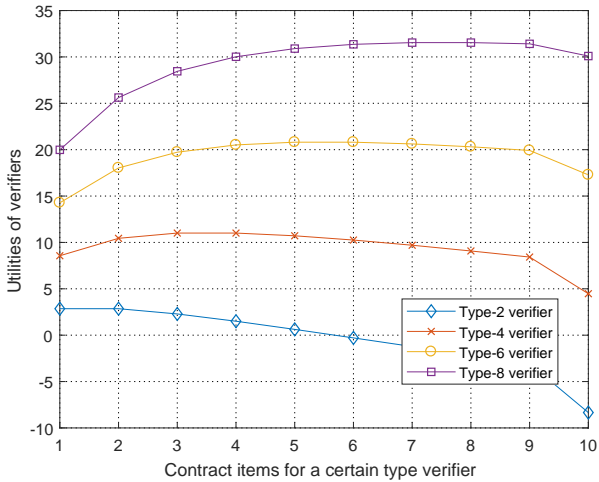}
  \caption{Utilities of verifiers under different contract items.}
   \label{VerifierUtility}
\end{figure}

\begin{figure}[t]\centering
  \includegraphics[width=0.5\textwidth]{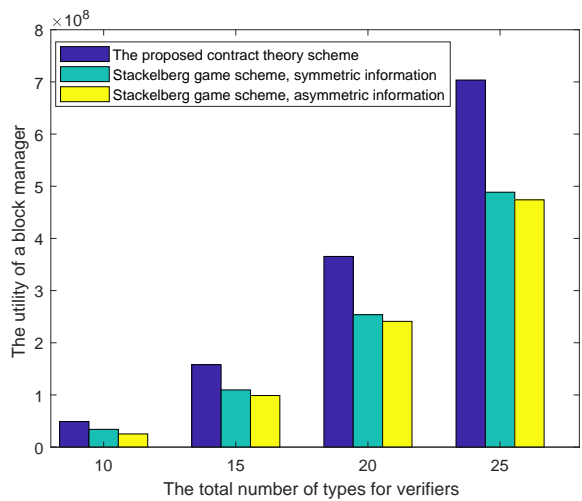}
  \caption{The utility of a block manager under different total number of verifier types.}
   \label{UbmUtility}
\end{figure}

Fig.~\ref{reputation_changed1} shows reputation variation of a malicious miner candidate  from the perspective of a well-behaved vehicles under three cases: (i) traditional DPoS scheme without reputation, (ii) TSL scheme, and (iii) MWSL scheme. In the traditional DPoS scheme without reputation, the reputation value of the compromised candidate evaluated by the vehicle is linear increasing because the well-behaved vehicle cannot detect the candidate's misbehaviors for other well-behaved vehicles. However, in the cases of TSL and our MWSL schemes, the reputation values of the candidate sharply decrease because of recommended opinions from other vehicles. The reputation value decreasing below reputation threshold of trusted miner in the MWSL scheme is faster than that of TSL because of the weights of interaction frequency, timeliness, and interaction effects on both recommended opinions and local opinions. This can avoid being misleading by compromised vehicles' recommended reputation opinions. As a result, our MWSL scheme achieves more accurate reputation calculation, and this therefore leads to more secure miner voting.

We observe the detection rate of 10 malicious miner candidates using the TSL and MWSL schemes during 60 minutes. Figure~\ref{detectionRate} shows that the MWSL scheme has much higher successful detection rate of malicious miners than that in the TSL scheme. We define a metric as the reputation threshold of successful detection, in which only the reputation of malicious miners below the threshold can be detected successfully.
When the reputation threshold of successful detection is 0.5,  the detection rate of the MWSL scheme is 100\%, which is 100\% higher than that of the TSL scheme. Due to higher detection rate in the MWSL scheme,  potential security threats can be removed more effectively, which leads to a securer BIoV.


From Fig.~\ref{detectionRate}, we can observe that successful detection probability is not good enough when the reputation threshold of successful detection is very low, e.g., 0.2. In the cases with a very low threshold, the active  miners generated by reputation voting may launch the verification collusion attack, that more than 1/3 active miners collude to generate false verification result for a data block \cite{chitchyan2018review,minercollusion}. To defend this intractable attack, standby miners should  participate in block verification to improve the correct probability of verified block. The correct probability of verified block means that the data block is correctly verified without the effects of the verification collusion attack. Figure~\ref{correctedProbability} shows the correct probability of data block after verification with respect to different reputation thresholds of successful detection. When the reputation threshold of successful detection is 0.2, the correct probability in our MWSL scheme with standby miners is 13\% higher than that of MWSL scheme without standby miners, while the TSL scheme without standby miners cannot defend against this collusion attack. This indicates that the proposed MWSL can ensure a secure block verification, even when attackers launch internal active miner collusion.

\subsection{Performance of  the incentive mechanism based on contract theory scheme}
A block manager acting as the contract publisher announces the designed contract items  to other active miners and standby miners. These miners choose a contract item ($R_q, L_q^{-1}$) to sign, and work as verifiers to finish the block verification task according to latency requirements in the signed contract. Finally, the verifiers obtain the corresponding incentives from the contract publisher. Figure~\ref{VerifierUtility} shows the utilities of verifiers with type 2, type 4, type 6 and type 8.
We can see that each type of verifiers obtains the maximum utility while selecting the contract item exactly designed for its type, which explains the IC constraint. All types of verifiers choose the contract items corresponding to their types with non-negative utilities, which validates the IR constraint \cite{zeng2018icc}.

We compare the profit of a block manager obtained from the proposed contract model, and Stackelberg game model from~\cite{8239591}. Figure~\ref{UbmUtility} shows that the profit of a block manager increases with the total number of verifier types. The more verifier types bring both more verifiers and contract item choices for high-type (high-reputation) verifiers, leading to the more secure block verification. The utility of the proposed contract model has better performance than that of the Stackelberg game model. The reason is that in the monopoly market, the proposed contract model provides limited contract items to extract more benefits from the verifiers. However, in the Stackelberg game model, rational  verifiers can optimize their individual utilities thus leading to less profit of the block manager. Moreover, the Stackelberg game model with symmetric information has better performance than that of Stackelberg game model with asymmetric information. The reason is that the game leader (the block manager) in the Stackelberg game with symmetric information can optimize its profit because of knowing the actions of  followers (verifiers), i.e., the symmetric information, and set the utilities of the follows as zero \cite{8239591}.

\section{Conclusion}
In this paper, we have introduced blockchain-based Internet of vehicles for secure P2P vehicle data sharing by using a hard security solution, i.e., the enhanced Delegated Proof-of-Stake consensus scheme. This DPoS consensus scheme has been improved by a two-stage soft security enhancement solution. The first stage is to select miners by reputation-based voting. A multi-weight subjective logic scheme has been  utilized to calculate securely and accurately the reputation of miner candidates. The second stage is to incentivize standby miners to participate in block verification using contract theory, which can further prevent internal collusion of active miners. Numerical results have indicated that our multi-weight subjective logic scheme has great advantages over traditional reputation  schemes in  improving detection rate of malicious miner candidates. Likewise, the proposed contract-based block verification scheme can further decrease active miners collusion and optimize the utilities of both the block manager and verifiers to further improve the security of vehicle data sharing. In the future work, we can further improve the accuracy of the miner candidates' reputation calculation through taking more weights into consideration.

\end{document}